\documentclass[%
 aip,
 amsmath,amssymb,
 reprint,%
]{revtex4-1}
\usepackage{ulem}
\usepackage{graphicx}
\usepackage{dcolumn}
\usepackage{bm}
\usepackage{color}
\usepackage{xcolor}
\usepackage[utf8]{inputenc}
\usepackage[T1]{fontenc}
\usepackage{mathptmx}
\usepackage{etoolbox}
\usepackage{tabularx}
\usepackage{lineno}
\usepackage{soul}

\newcommand{\CsVSb}{{CsV$_{3}$Sb$_{5}$}}
\newcommand{\AVSb}{{AV$_{3}$Sb$_{5}$}} 

\newcommand{\RbVSb}{{RbV$_{3}$Sb$_{5}$}}
\newcommand{\pll}{\kern 0.56em/\kern -0.8em /\kern 0.56em}

\makeatletter
\def\@email#1#2{%
 \endgroup
 \patchcmd{\titleblock@produce}
  {\frontmatter@RRAPformat}
  {\frontmatter@RRAPformat{\produce@RRAP{*#1\href{mailto:#2}{#2}}}\frontmatter@RRAPformat}
  {}{}
}%
\makeatother
\begin{document}


\title[]{Superconducting critical temperature and dimensionality tuning of RbV$_3$Sb$_5$ via biaxial strain}
\author{Tsz~Fung~Poon$^\S$}
\author{King~Yau~Yip$^\S$}
\affiliation{Department of Physics, The Chinese University of Hong Kong, Shatin, Hong Kong, China}
\author{Ying~Kit~Tsui}
\affiliation{Department of Physics, The Chinese University of Hong Kong, Shatin, Hong Kong, China}
\affiliation{Quantum Science Center of Guangdong-Hong Kong-Macao Greater Bay Area, Shenzhen, China}
\author{Lingfei~Wang}
\author{Kai~Ham~Yu}
\author{Wei~Zhang}
\author{Zheyu~Wang}
\affiliation{Department of Physics, The Chinese University of Hong Kong, Shatin, Hong Kong, China}
\author{Taketo~Nakatani}
\author{Chishiro~Michioka}
\affiliation{Division of Chemistry, Graduate School of Science, Kyoto University, Kyoto, Japan}
\author{Hiroaki~Ueda}
\affiliation{Co-Creation Institute for Advanced Materials, Shimane University, Matsue, Japan}
\author{Siu~Tung~Lam}
\affiliation{Department of Physics, The Chinese University of Hong Kong, Shatin, Hong Kong, China}
\author{Kwing~To~Lai}
\affiliation{Department of Physics, The Chinese University of Hong Kong, Shatin, Hong Kong, China} 
\affiliation{Shenzhen Research Institute, The Chinese University of Hong Kong, Shatin, Hong Kong, China}
\author{Swee~K.~Goh$^*$}
\email[]{skgoh@cuhk.edu.hk}
\affiliation{Department of Physics, The Chinese University of Hong Kong, Shatin, Hong Kong, China}

\date{\today}

\begin{abstract}
Kagome metal \AVSb~(A=K, Rb, Cs) has emerged as an intriguing platform for exploring the interplay between superconductivity and other quantum states. Among the three compounds, \RbVSb\ has a notably lower superconducting critical temperature ($T_c$) at ambient pressure, posing challenges in exploring the superconducting state. For instance, the upper critical field ($H_{c2}$) is small and thus difficult to measure accurately against other control parameters. Hence, enhancing superconductivity would facilitate $H_{c2}$ measurements, providing insights into key superconducting properties such as the dimensionality. In this letter, we report the tuning of the $T_c$ in \RbVSb\ through the application of biaxial strain. Utilizing a negative thermal expansion material ZrW$_2$O$_8$ as a substrate, we achieve a substantial biaxial strain of $\epsilon=1.50\%$, resulting in a remarkable 75\% enhancement in $T_c$. We investigate the $H_{c2}$ as a function of temperature, revealing a transition from multi-band to single-band superconductivity with increasing tensile strain. Additionally, we study the $H_{c2}$ as a function of field angle, revealing a plausible correlation between the $T_c$ enhancement and the change in dimensionality of the superconductivity under tensile strain. Further analysis quantitatively illustrates a transition towards two-dimensional superconductivity in \RbVSb\ when subjected to tensile strain. Our work demonstrates that the application of biaxial strain allows for the tuning of both the $T_c$ and superconducting dimensionality in \RbVSb.

\end{abstract}

\maketitle
The quest for enhanced superconductivity remains one of the foremost endeavors in the field of condensed matter physics. When a superconductor has a low critical temperature ($T_c$), the associated critical magnetic fields are typically small, hindering the examination of the superconducting state through the measurement of the critical fields. Efforts to tune $T_c$ include the application of hydrostatic pressure, which shrinks the crystal volume isotropically and modifies both the electronic band structure and the phonon spectrum. Successful examples of a significant increase in $T_c$ by hydrostatic pressure include iron-based~\cite{Takahashi2008, Kotegawa2009,Alireza2008, Goh2010,Yip2017} and cuprate~\cite{Markert1989, Gao1994,
 Goldschmidt1996,Monteverde2005} superconductors. Superconductors with reduced crystal dimensionality, such as two-dimensional (2D) or quasi-2D compounds, have been heavily studied in recent years. The ability to tune the superconductivity of these systems is similarly important. Analogous to the case of tuning three-dimensional bulk superconductors, biaxial strain, which can vary the crystal area isotropically, emerges as a promising tuning tool for studying 2D superconductors.

A simple biaxial device can be assembled by mechanically bonding the superconductor to a substrate with different thermal expansion coefficients. Due to this thermal expansion mismatch between the sample and the substrate, a biaxial strain develops in-plane during cooling and reaches saturation at the zero-temperature limit. This novel technique has been utilized for the studies of superconductivity in various materials, such as those reported in Refs.~[\onlinecite{Nakajima2021, Bohmer2017, Yip2023, Yip2024}]. Given the simplicity of the setup, it can be conveniently accommodated in various sample spaces. For example, the biaxial device can be rotated in most laboratory magnets, enabling the measurement of the upper critical field ($H_{c2}$) as a function of the field angle. 

Recently, the kagome metal \AVSb\ (A=K, Rb, Cs)~\cite{Ortiz2019, Ortiz2020} has emerged as an intriguing platform for exploring the interplay among its various quantum states. This is because the kagome lattice intrinsically hosts flat bands, van Hove singularities, and Dirac points in the electronic structure~\cite{Ortiz2019, Ortiz2020, Neupert2022, Li2021, Jiang2023, Yin2022}. Within the three members in the \AVSb\ family, \CsVSb\ is the most heavily studied, arguably because it has the highest $T_c$ ($\sim$2.7~K) at ambient pressure~\cite{Chen2021a, Zhao2021, Yu2021b, Yu2021, Wang2021charge, Han2023, Kang2023a,Zhang2023,Zhang2024}. The lower $T_c$ members,  KV$_3$Sb$_5$~\cite{Du2021,Yang2020,Luo2022, Wang2023b, Wang2025, Guguchia2023, Zhu2022} and \RbVSb ~\cite{Zhu2022, Guguchia2023, Wang2023, Yin2021,Du2022, Guguchia2023, Wang2021a}, deserve more experimental attention. Thus, it is desirable to increase their $T_c$ systematically.

In this manuscript, we apply the biaxial strain tuning methodology to investigate \RbVSb. We first report the superconducting transition in \RbVSb\ coupled to different substrates. 
For the device with the enhanced $T_c$, we further measure $H_{c2}$ against the temperature as well as the field angle. Our data allow us to identify the driving force behind the $T_c$ enhancement, and discuss the dimensionality of the superconductivity. We also demonstrate how a large tensile strain can be achieved, by resorting to a negative thermal expansion material (ZrW$_2$O$_8$)~\cite{Evans1996}. Our results showcase the versatility of the biaxial strain method and pave the way to the understanding of the superconductivity in the \AVSb\ family.

High-quality single crystals of \RbVSb\ were synthesized by the self-flux method as described in Ref.~[\onlinecite{Wang2023}]. Single crystals from the same batch were cleaved and cut into small strips of typical dimensions 800~$\times$~250~$\times$~10~$\mu$m$^3$ for measurements under biaxial strain $\epsilon$. The ZrW$_2$O$_8$ substrates were prepared from ZrO$_2$ (Nacalai Tesque, 98\%) and WO$_3$
(Nacalai Tesque, 99.5\%) powders. These powders were mixed stoichiometrically, pressed into pellets, and heated for 12 hours at 1200$^\circ$C, the temperature determined based on the phase diagram given by Ref.~[\onlinecite{Chang1967}]. The pellets were then quenched in liquid nitrogen to prevent decomposition into ZrO$_2$ and WO$_3$. The pellets were subsequently ground, pressed into thin disks, heated, and quenched again. X-ray diffraction data, presented in Supplementary Material, confirms the formation of ZrW$_2$O$_8$. Biaxial strain can be induced on the sample by the difference in thermal expansion between the sample and the substrate. The small size of the sample ensures that strain is effectively imposed on the sample by the substrate. Compressive strain ($\epsilon<0$) was induced by attaching \RbVSb\ to polycarbonate substrates with Cyanoacrylate (CN) adhesives. Similarly, different magnitudes of the tensile strain ($\epsilon>0$) were induced by attaching \RbVSb\ to cover glass, sapphire, diamond, and ZrW$_2$O$_8$ substrates with ThreeBond 2086M two-component epoxy resin adhesive (ThreeBond Holdings Co., Ltd.). In the process of attaching the samples to these substrates, the substrates were slightly heated for the better flow of the adhesive. The strain of sample and that of substrate were characterized by strain gauges (Tokyo Measuring Instruments Lab Co.,~Ltd.). The resultant strain induced on the sample at zero-temperature limit can be calculated as described in Ref.~[\onlinecite{Yip2024}]. See the Supplementary Material for more details. A standard four-probe method was used to measure electrical resistance. A Bluefors dilution refrigerator provides the low-temperature environment down to 10~mK. A superconducting vector magnet was adopted for applying a vector field of 3~T/5~T ($y$/$z$-axes) to study the angular dependence of the upper critical field.

\begin{figure}[!t]\centering
      \resizebox{9cm}{!}{
              \includegraphics{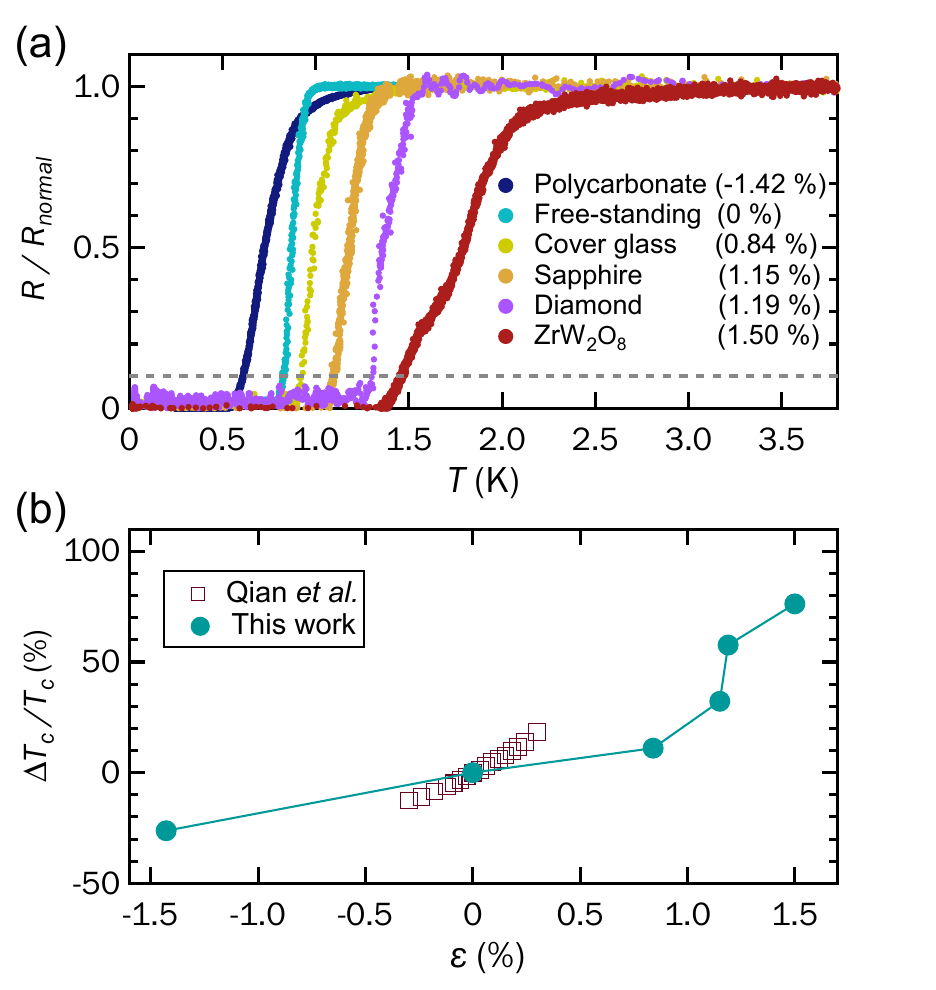}}\
              \caption{\label{fig1} 
              (a) Temperature dependence of normalized resistance curves, showing the superconducting transitions of \RbVSb\ with different substrates. The calculated strain induced in the low-temperature limit is shown in the brackets. The horizontal gray dashed line indicates the $10\%$ criterion for the definition of $T_c$. (b) Percentage change of superconducting temperature $\Delta T_c/T_c$ as the function of strain summarized from (a), the square marker represents the $\epsilon_{A\rm{_{1g}}}$ component extracted from the uniaxial strain study on \CsVSb\ from Qian $et$ $al$.~\cite{Qian2021}. 
              }
             
\end{figure}


Figure~\ref{fig1}(a) shows the normalized temperature dependence of electrical resistance around the superconducting transition for \RbVSb\ experiencing different strains. For the free-standing sample ($\epsilon=0$), $T_c$ is 0.83~K. Here, $T_c$ is defined as the temperature at which the resistance drops to 10$\%$ of the normal state value. Under compressive strain, $T_c$ decreases to 0.61~K. Thus, in an attempt to increase $T_c$, we look for materials with a thermal expansion coefficient smaller than that of \RbVSb\ to induce a tensile strain in the sample. For example, diamond is known to have a negligible thermal expansion coefficient: at 4~K, the lattice parameter $a$ only decreases by $3\times10^{-4}$\AA\ from the room temperature value of 3.5671~\AA~\cite{Stoupin2010}. As shown in Fig.~\ref{fig1}(a), $T_c$ of \RbVSb\ placed on diamond is enhanced to 1.30~K. Enhancement in $T_c$ has also been observed when sapphire and cover glass were used. To create an even larger tensile strain, we turned to ZrW$_2$O$_8$, which is a material with a negative thermal expansion~\cite{Evans1996}.  The usage of ZrW$_2$O$_8$ substrate allows us to achieve a large tensile strain of $\epsilon=1.50 \%$. Correspondingly, $T_c$ is enhanced to a value as large as 1.46~K. 

Figure~\ref{fig1}(b) shows the percentage change in $T_c$ ($\Delta T_c/T_c$) extracted from our result for \RbVSb\ (filled circles) and that from Ref.~[\onlinecite{Qian2021}] for \CsVSb\ (open squares). Here, $\Delta T_c/T_c$ is defined as $[T_c(\epsilon)-T_c(\epsilon=0)]/T_c(\epsilon=0)$.  For a valid comparison, we extract the biaxial component from the uniaxial strain experiment described in Ref.~[\onlinecite{Qian2021}]. Specifically, we are interested in the isotropic response $\epsilon_{A\rm{_{1g}}}$ of the uniaxial strain experiment, as described in Supplementary Material. For \RbVSb, $\Delta T_c/T_c$  increases approximately linearly as $\epsilon$ increases from $-1.42\%$ to $0.84\%$. This is followed by an upturn when $\epsilon$ exceeds $0.84\%$, with $\Delta T_c/T_c$ reaching $75\%$ when $\epsilon=1.50\%$. On the other hand, $\Delta T_c/T_c$ as the function of (biaxial) strain in \CsVSb\ is dominated by the linear term without slope changing. Overall, the use of substrates with progressively smaller, and even negative, thermal expansion allowed us to exert large tensile strain on \RbVSb, revealing a surprising upturn in $\Delta T_c/T_c$. It would be interesting to explore if the upturn can be detected in \CsVSb\ with a larger tensile strain.  Typically, due to the Poisson effect, when a tensile biaxial strain is applied in the $ab$-plane, a compressive strain is induced along the $c$-axis. Similarly, when a tensile uniaxial strain is applied along an in-plane axis, a compressive strain is induced along the orthogonal in-plane axis and the $c$-axis. In Ref.~[\onlinecite{Qian2021}], the strain along the $c$-axis is argued to be the dominant driving factor for $T_c$ enhancement. This viewpoint is further supported by other strain studies on \CsVSb~\cite{Yang2023,Frachet2024}. Since both \RbVSb\ and \CsVSb\ respond to strain in a similar manner, as suggested by the similar trend of $T_c$ enhancement, the underlying tuning parameter of $T_c$ for \RbVSb\ is also the strain along the $c$-axis, and the strain in the $ab$-plane only plays a marginal effect. All in all, the strain along the $c$-axis appears to be the main driving force in $T_c$ enhancement in \RbVSb\ and \CsVSb.

\begin{figure}[!t]\centering
       \resizebox{9cm}{!}{
              \includegraphics{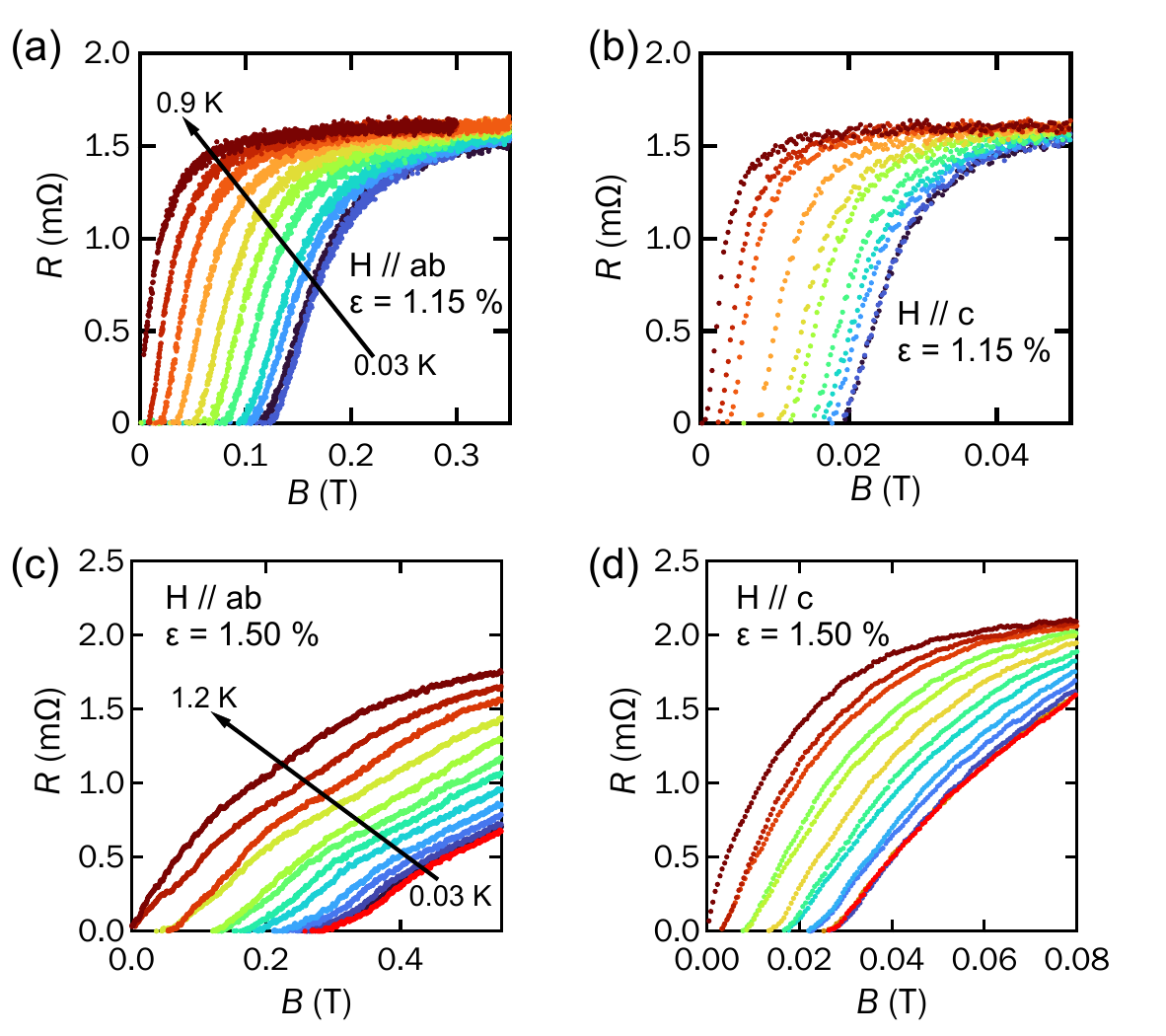}}  
              
              \caption{\label{fig2} 
             Field dependence of the electrical resistance in \RbVSb\ under tensile strain at different temperatures, for (a)~$\epsilon=1.15\%$ and $H\pll ab$, (b)~~$\epsilon=1.15\%$ and $H\pll c$, (c)~~$\epsilon=1.50\%$ and $H\pll ab$, and (d)~~$\epsilon=1.50\%$ and $H\pll c$.}
\end{figure}

The largely enhanced $T_c$ in \RbVSb\ under biaxial tensile strain facilitates the measurement of the field dependence of the electrical resistance for the extraction of the $H_{c2}$.
Figures~\ref{fig2}(a) and \ref{fig2}(b) show the field dependence of the resistance in \RbVSb\ under tensile strain of $\epsilon=1.15\%$, with the applied field in the $ab$-plane~($H\pll ab$) and along the $c$-axis~($H\pll c$), respectively. Likewise, Figs.~\ref{fig2}(c) and \ref{fig2}(d) show the field dependence of the resistance in \RbVSb\ under the largest tensile strain of $\epsilon=1.50\%$ for the two field orientations. In all cases, the zero resistance persists to higher field as the temperature decreases. To understand the pair-breaking mechanism of Cooper pairs associated with the drastic enhancement in $T_c$ of \RbVSb\ under tensile strain, we examine the  $H_{c2}$ as a function of temperature. 

\begin{figure}[!t]\centering
       \resizebox{9cm}{!}{
              \includegraphics{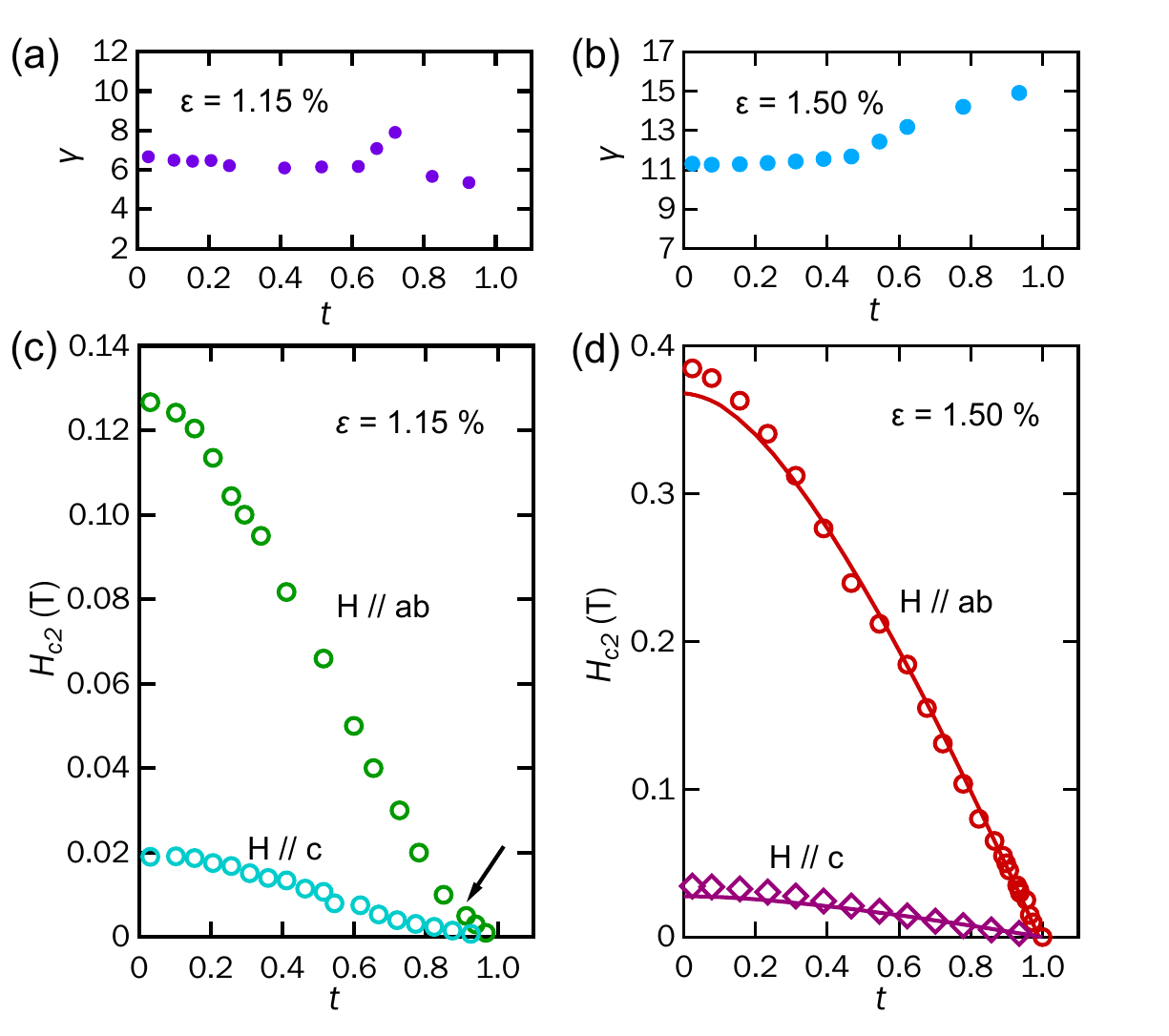}}                				
              \caption{\label{fig3} Temperature dependence of upper critical field $H_{c2}(t)$ in \RbVSb\ under tensile strain of (c)~$\epsilon=1.15\%$ and (d)~$\epsilon=1.50\%$, where $t=T/T_c$ is the reduced temperature. The solid lines in (d) represent the simulated $H_{c2}(t)$ by the WHH model. Temperature dependence of the extracted $\gamma$ for (a)~$\epsilon=1.15\%$ and (b)~$\epsilon=1.50\%$.}
\end{figure}

Figures~\ref{fig3}(c) and \ref{fig3}(d) show the temperature dependence of the $H_{c2}$ in \RbVSb\ under tensile strain of $\epsilon=1.15\%$ and $\epsilon=1.50\%$, respectively. Here, $H_{c2}$ is defined as the magnetic field at which the resistance drops to 10\% of that in the zero-field normal state, and the temperature axis is expressed in the reduced unit, namely $t=T/T_c$. For $\epsilon=1.15\%$, $H_{c2}(t)$ shows trace of an upturn~(indicated by the black arrow in ~Fig. 3(c)) near $t=1$ for $H\pll ab$. Such an upturn feature has also been reported for \CsVSb, and it has been interpreted as a sign of multi-band superconductivity~\cite{Ni2021}. Surprisingly, when the tensile strain increases to $\epsilon=1.50\%$, the upturn disappears, suggesting single-band superconductivity~\cite{Werthamer1966}. In addition, the $H_{c2}(t)$ shows a linear dependence over a wide temperature range from $t=1$. Therefore, we simulate the single-band Werthamer–Helfand–Hohenberg~(WHH) model~\cite{Werthamer1966}, with Maki parameter $\alpha=0$ and the spin-orbit coupling $\lambda=0$ for both field directions. The results of the simulations are represented by the solid lines, which are in good agreement with the experimental data (Fig.~\ref{fig3}(d)). The orbital-limited field at zero temperature $H^{orb}_{c2}(0)$ can be obtained from the WHH simulations, and they are found to be found to be 0.37~T and 0.03~T for $H\pll ab$ and $H\pll c$, respectively. Overall, the results imply that upon a moderate tensile strain ($\epsilon=1.15\%$), \RbVSb\ exhibit the multi-band behavior, which is commonly shared in the \AVSb\ family~\cite{Ni2021}. By applying an unprecedentedly large tensile strain on \RbVSb, we observed a single band behavior in $H_{c2}(t)$.

Next, we investigate the anisotropy factor $\gamma$, defined as $\gamma = H^{||ab}_{c2}/H^{||c}_{c2}$, where the superscript stands for the field direction. Figures~\ref{fig3}(a) and (b) show the temperature dependence of $\gamma$ for the tensile-strained \RbVSb\ with $\epsilon=1.15\%$ and $\epsilon=1.50\%$, respectively. Comparing the two cases, with the larger tensile strain, $\gamma$ is significantly larger for the whole temperature range. In particular, at 30~mK, $\gamma$ is 6.7 for $\epsilon=1.15\%$ but it is significantly enhanced to 11.3 for $\epsilon =1.50\%$. Therefore, our results show that the superconducting anisotropy is sensitive to the applied biaxial strain. In addition, for the two levels of tensile strain, $\gamma$ shows clear enhancement with temperature at around $t=0.5$, with the enhancement persists to $t\rightarrow1$ for $\epsilon=1.50\%$, but not for $\epsilon=1.15\%$. Such a difference may be related to the difference in the dimensionality of the superconductivity. For example, $\gamma$ diverges when $t\rightarrow 1$ for 2D superconductivity~\cite{Tinkham1963}. The differences in the magnitudes of $\gamma$ for two tensile-strained \RbVSb\ samples and their distinct responses to temperature suggest that the superconducting dimensionality and the enhancement of $T_c$ under tensile strain may be correlated.

\begin{figure}[!t]\centering
       \resizebox{9cm}{!}{
              \includegraphics{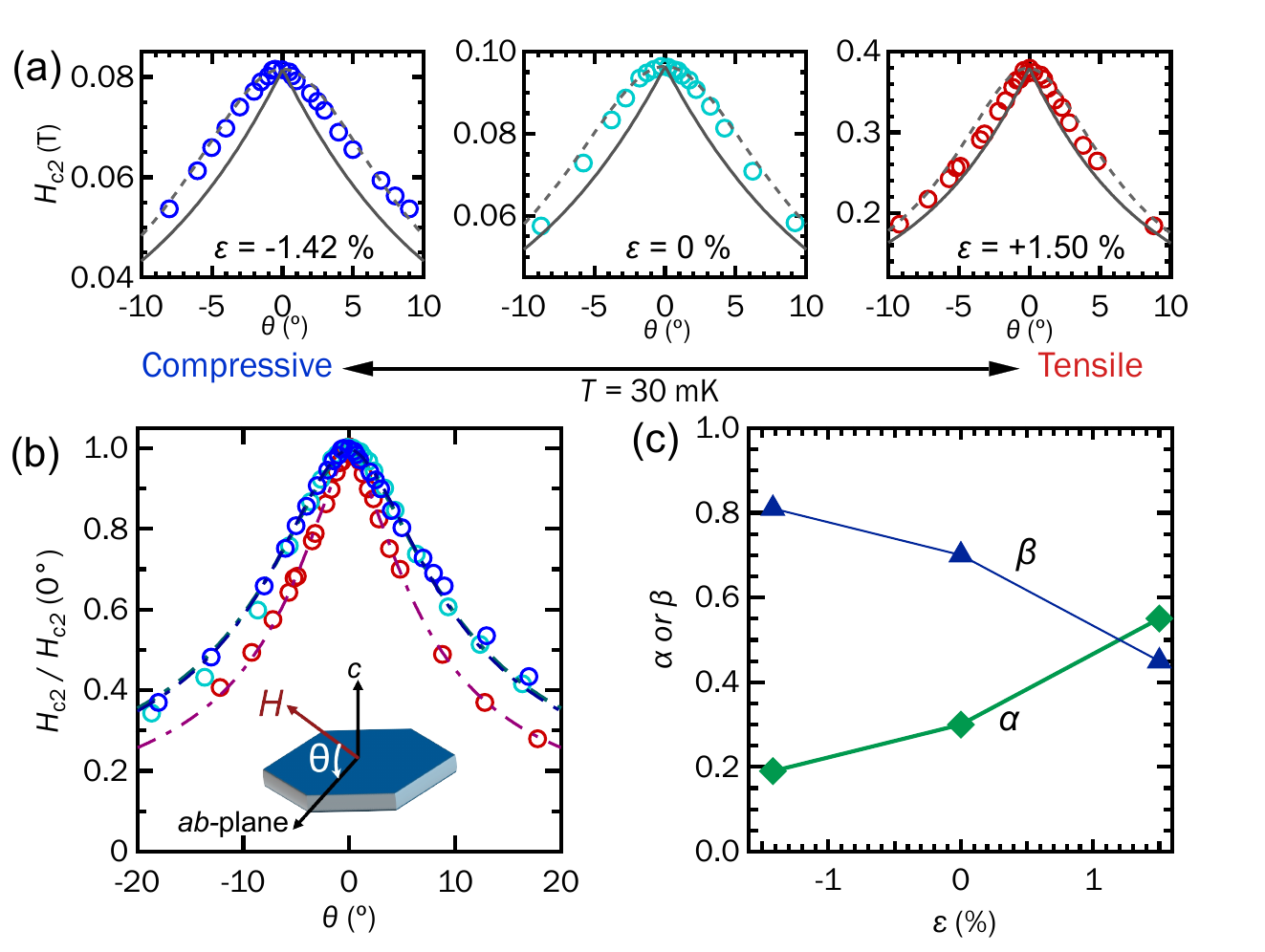}}                				
              \caption{\label{fig4} (a) Angular dependence of upper critical field ($H_{c2}(\theta)$) near $\theta=0^{\circ }$ with different strains at 30~mK, the solid line represents the simulated $H_{c2}(\theta)$ by Tinkham model while the dash gray line represents the simulated $H_{c2}(\theta)$ by Ginzburg-Landau anisotropic mass model. (b) Normalized $H_{c2}(\theta)$ near $\theta=0^{\circ }$ with different strains at 30~mK, fitted with Eqn.~\ref{equ1}. (c) The fitting coefficients $\alpha$ and $\beta$ as the function of strain.}
\end{figure}

To get a deeper understanding, we directly probe the superconducting dimensionality by measuring the angular dependence of the upper critical field ($H_{c2}(\theta)$). The schematic of the sample orientation and field direction is shown in the inset of Fig.~\ref{fig4}(b). Figure~\ref{fig4}(a) shows the $H_{c2}(\theta)$ under different levels of strain at 30~mK. Here, we use the anisotropic mass Ginzburg-Landau~model~(GL model) and the Tinkham model~\cite{Tinkham1963}, which are commonly employed to describe 3D and 2D superconductivity, respectively. Their major difference is that, $H_{c2}(\theta)$ is smooth for all $\theta$ in the GL model, but a cusp appears at $\theta=0^{\circ }$ in the Tinkham model. We simulate $H_{c2}(\theta)$ using both the GL model and the Tinkham model for the three cases shown in Fig.~\ref{fig4}(a) and present them as dotted line and solid line, respectively. For these simulations, we have used the experimental $H_{c2}(0^\circ)$ and $H_{c2}(90^\circ)$ at respective strains as input parameters. For the compressive strain~($\epsilon = -1.42\%$) and free-standing~($\epsilon = 0\%$) cases, $H_{c2}(\theta)$ displays a smooth maximum across $\theta=0^{\circ}$, and the datasets are better described by the GL model. For the tensile strain case, the $H_{c2}(\theta)$ exhibits a sharper maximum across $\theta=0^{\circ}$, and the dataset falls between the simulations by the GL and Tinkham model. Therefore, the simulations imply a change in dimensionality of the superconductivity in \RbVSb\ under tensile strain.  
To better fit the dataset and quantify the contribution of 2D and 3D superconductivity involved in the $H_{c2}(\theta)$ of \RbVSb\ under different levels of strain, we adopt the combination of GL model and Tinkham model~\cite{Goh2012, Shimozawa2014,Uchida2019, Chow2023}, which have the following form:
\begin{equation}
\left[\dfrac{H_{c2}(\theta)\cos\theta}{H_{c2}(0^{\circ })}\right]^2+
\alpha\left|\dfrac{H_{c2}(\theta)\sin\theta}{H_{c2}(90^{\circ })}\right|+\beta\left[\dfrac{H_{c2}(\theta)\sin\theta}{H_{c2}(90^{\circ})}\right]^2=1 
\label{equ1}
\end{equation}
where $\alpha$ and $\beta$ are the fitting parameters that quantify how closely $H_{c2}(\theta)$ resembles the Tinkham model and the GL model respectively. We anticipate the sum of $\alpha$ and $\beta$ to be 1, and this is confirmed by our fitting results. Figure~\ref{fig4}(b) shows the fits of normalized $H_{c2}(\theta)$ at the different levels of strain to the combined model. All the datasets can be well described by the combined model. The extracted $\alpha$ and $\beta$ for different strain values are shown in Fig.~\ref{fig4}(c). For the free-standing case, $\alpha$ is $0.30$, which indicates a larger contribution of 3D superconductivity. Upon applying a compressive strain, the $\alpha$ is $0.19$, indicating that the 2D character is suppressed, approaching the GL model~(3D) limit. On the other hand, with tensile strain, $\alpha$ increases to $0.55$, indicating a shift in the dimensionality towards the 2D side. 
Interestingly, under the free-standing condition, \CsVSb\ already exhibits 2D superconductivity~\cite{Hossain2025}, and its $T_c$ is higher than that of \RbVSb. Since the $T_c$ enhancement in \RbVSb\ is accompanied by the shift in the dimensionality of the superconductivity from predominantly 3D to a more 2D character, the results suggests that 2D superconductivity plays an important role in the $T_c$ enhancement in the \AVSb\ family.

While hydrostatic pressure has previously been employed to enhance $T_c$ of \RbVSb~\cite{Wang2021a}, the present study provides a different yet complementary avenue for tuning the superconductivity in \RbVSb. We argue that the methodology used herein has several distinct advantages. First, hydrostatic pressure is inherently compressive, whereas biaxial strain can be either compressive or tensile in the planar direction. This unique property of biaxial strain allows for greater flexibility in tuning. The fact that only two spatial directions are actively modified further enables us to conclude that the $c$-axis length serves as the decisive tuning parameter for $T_c$ in \RbVSb. Second, the biaxial device offers an excellent platform for optical measurements, because the sample is not obstructed, unlike the case of hydrostatic pressure where the optical path is limited by the materials surrounding the samples. Thus, the investigation of strained \RbVSb\ with optical methods can be suitable for further projects.

The shift towards a superconducting state with a more 2D character could be due to a more 2D-like Fermi surface  under tensile strain. This trend might be counterintuitive, but it is possible that the electronic structure has been modified in such a way that some 3D Fermi pockets disappear when the $c$-axis is shorter, which happens when an in-plane tensile biaxial strain is applied. This would allow the 2D character to dominate, as reflected in the dimensionality of superconductivity. This unusual behavior has been reported in MoTe$_2$, where the Fermi surface becomes more 2D-like under pressure~\cite{Hu2020}. Furthermore,  a Lifshitz transition can be associated with the disappearance of 3D pockets, which could explain the observed jump in $\Delta T_c/T_c$ displayed in Fig.~\ref{fig1}(b). Quantum oscillation data against the field angle can probe the dimensionality of the Fermi surface. The current design is compatible with such a measurement, allowing the discussion of both the superconducting and the normal state anisotropies by conducting measurements in strong magnetic fields. Recently, we demonstrated the measurement of quantum oscillations up to 60~T in a pulsed magnet~\cite{Yip2024}. Thus, we do not anticipate significant barriers in measuring quantum oscillations in biaxially strained \RbVSb\ against the field angle.

To summarize, we have studied the biaxial strain effect on the superconductivity in \RbVSb, using substrates with different thermal expansion coefficients. In particular, $T_c$ is enhanced drastically from 0.82~K to 1.46~K with the maximum tensile strain ($\epsilon=1.50\%$) exerted by the negative thermal expansion substrate ZrW$_2$O$_8$. We further examine the temperature dependence of $H_{c2}$ in tensile strained \RbVSb. Multi-band behavior is inferred for $\epsilon=1.15\%$, while single-band behavior arises for $\epsilon=1.50\%$. The significant increase in the extracted anisotropy factor $\gamma$ suggests the correlation between $T_c$ enhancement and the shift in the superconducting dimensionality. This shift is quantified by the model that combines the Tinkham model and the anisotropic mass Ginzburg-Landau model, showing the shift towards 2D superconductivity. Our work consolidates the biaxial strain as a promising tool to tune quasi-2D materials, and paves the way for the understanding of the superconductivity in \AVSb\ family.

\section*{supplementary material}
The X-ray diffraction pattern of ZrW$_2$O$_8$, nominal strain induced on \RbVSb\ by substrates, and extraction of isotropic response $\epsilon_{A\rm{_{1g}}}$ from uniaxial strain experiment are described in the Supplementary Material.

\begin{acknowledgments}
$^\S$T. F. Poon and K. Y. Yip contributed equally to this work. The work was supported by the Research Grants Council of Hong Kong (Grant Nos. A-CUHK 402/19, CUHK 14301020, CUHK 14300722, CUHK 14302724), CUHK Direct Grant (4053577, 4053664), and the Guangdong Provincial Quantum Science Strategic Initiative (Grant No. GDZX2301009).
\end{acknowledgments}

\section*{AUTHOR DECLARATIONS
}
\subsection*{{Conflict of Interest
}}
The authors have no conflicts to disclose.

\section*{Data Availability Statement}
The data that support the findings of
this study are available from the
corresponding author upon reasonable
request.

\providecommand{\noopsort}[1]{}\providecommand{\singleletter}[1]{#1}%

\end{document}